\documentclass[singlecolumn,preprintnumbers,amsmath,amssymb,notitlepage,nofootinbib,longbibliography,superscriptaddress]{revtex4-1}

\usepackage{natbib}
\usepackage{graphicx}
\usepackage{dcolumn}
\usepackage{dsfont}
\usepackage{bm}
\usepackage{times}
\usepackage{ulem}

\usepackage{xcolor}
\usepackage{url}
\usepackage[colorlinks=true,breaklinks=true,allcolors=blue]{hyperref}
\usepackage{microtype}

\newcommand{\RR}{\mathbb{R}}

\newcommand{\proof}{\noindent\textbf{Proof:} }
\newcommand{\proofend}{$\blacksquare$}

\definecolor{jens}{rgb}{0,.7,.3}
\newcommand{\je}[1]{{\color{jens} #1}}

\begin{document}

\title{Lieb-Robinson bounds for open quantum systems with long-ranged interactions}

\author{Ryan Sweke}
\affiliation{\mbox{Dahlem Center for Complex Quantum Systems, Freie Universit\"{a}t Berlin, 14195 Berlin, Germany}}

\author{Jens Eisert}
\affiliation{\mbox{Dahlem Center for Complex Quantum Systems, Freie Universit\"{a}t Berlin, 14195 Berlin, Germany}}

\author{Michael Kastner} 
\email{kastner@sun.ac.za} 
\affiliation{National Institute for Theoretical Physics (NITheP), Stellenbosch 7600, South Africa} 
\affiliation{\mbox{Institute of Theoretical Physics,  Department of Physics, University of Stellenbosch, Stellenbosch 7600, South Africa}}

\date{\today}

\begin{abstract}
We state and prove four types of Lieb-Robinson bounds valid for many-body open quantum systems with power law decaying interactions undergoing out of equilibrium dynamics. 
We also provide an introductory and self-contained discussion of the setting and tools necessary to prove these results. The results found here apply to physical systems in which both long-ranged interactions and dissipation  are present, as
commonly encountered in certain quantum simulators, such as
Rydberg systems or Coulomb crystals formed by ions.
\end{abstract}

\maketitle
 
\section{Introduction}

For many non-relativistic lattice models, despite the absence of a finite maximum propagation speed in the strict sense, it has been firmly established by now that physical effects are mostly restricted to a causal region in space time, with only small ``leakage'' into the region outside the causal region. The mathematical tools for stating and proving such a quasilocal structure go under the name of Lieb-Robinson bounds. These are usually
stated as upper bounds on the operator norm of commutators of the form
\begin{equation}\label{e:LR_general}
\left\lVert[A(t),B]\right\rVert\leq b(t,x),
\end{equation}
where $A$ and $B$ are observables, $A(t)$ denotes the operator $A$ time-evolved in the Heisenberg picture, and $x=d(A,B)$ is the spatial separation (usually with respect to the 1-norm) of the supports of $A=A(0)$ and $B$ on the lattice. In Lieb and Robinson's original work \cite{LiebRobinson72}, which is valid for systems on regular lattices with finite local Hilbert space dimension and finite-range interactions, a bound of the form
\begin{equation}\label{e:LRoriginal}
b_{\text{LR}}(t,x)=c\exp\left(\frac{vt-x}{\xi}\right)
\end{equation}
has been derived. The constants $c$, $v$, and $\xi$ depend on general features like lattice dimension and the interaction strength, but not on the details of the model. This bound shows that there is a region in the $(t,x)$-plane, outside the cone defined by $vt\geq x$, where the norm of $[A(t),B]$ is strongly suppressed and decays exponentially with the distance $x$.
	
While the commutator on the left-hand side of \eqref{e:LR_general} may not be of immediate physical interest in itself, it can be conveniently
used to derive bounds on a number of physically relevant quantities. These include the propagation in space and time of 2-point \cite{BravyiHastingsVerstraete06,Kastner15} and $n$-point correlations \cite{Tran_etal17}, of entanglement, and of quantum information \cite{BravyiHastingsVerstraete06}. Moreover, Lieb-Robinson bounds have been used to prove static properties, like the exponential clustering of correlations in ground states \cite{HastingsKoma06}, or a higher-dimensional version of the Lieb-Schulz-Mattis theorem \cite{Hastings04}. Known insights into the stability of quantum phases, and in particular the stability of topological order, also derive from Lieb-Robinson bounds \cite{TopologicalStability}.
	
Following Lieb and Robinson's original result for quantum systems on regular lattices with finite-range interactions, similar results have been obtained in different or more general settings, including quantum systems on general graphs (instead of regular lattices) \cite{NachtergaeleSims06}, models with long-ranged interactions \cite{HastingsKoma06,FossFeigGongClarkGorshkov15,StorchvandenWormKastner15,EisertvdWormManmanaKastner13}, disordered systems \cite{BurrellOsborne07,AbdulRahman_etal17}, open quantum systems \cite{Poulin10,KlieschGogolinEisert14, Barthel_OQSLR_2012, Kliesch_DQCT_2011,Nachtergaele_2011}, and classical lattice models \cite{Marchioro_etal78,MetivierBachelardKastner14,Matsuta2016}. While the obtained bounds may differ in their functional forms, they all have in common that they specify a certain (not necessarily cone-shaped) causal region, outside of which $\left\lVert[A(t),B]\right\rVert$ is smaller than some chosen $\epsilon>0$ and decays further away from that region (although not necessarily exponentially).
	
A class of systems for which Lieb-Robinson bounds have not been available so far is 
open quantum lattice models with power law decaying long-ranged interactions. It is the purpose of the present work  to fill this gap. Such systems have seen a lot of interest recently, especially due to their relevance for a number of experimental platforms that make use of trapped cold atoms, molecules, or ions. Often these platforms are employed for quantum simulation of unitary dynamics, but it turns out that significant non-unitary effects, like dissipation and decoherence, frequently have to be accounted for as well, or may even act as desirable resources. Examples of experimental realizations of open quantum lattice models with long-ranged interactions include Coulomb crystals of trapped ions \cite{FossFeig_etal13,Bohnet_etal16,Shankar_etal17,TrautmannHauke18}, lattices of Rydberg atoms \cite{Malossi_etal14,Schempp_etal15,Zeiher_etal16,SchoenleberBentleyEisfeld18,Browaeys}, and laser-driven atomic clouds \cite{Ott_etal13,GuerinAraujoKaiser16,Pucci_etal17}.
	
In this work we state and prove four different Lieb-Robinson bounds. The first three of these bounds are valid for open quantum lattice systems of Lindblad form, with finite local Hilbert space dimension at each lattice site, and for interactions whose strength decays in some suitable way with the graph distance $d(i,j)$ between lattice sites $i$ and $j$. While the setting we use applies to very general models on arbitrary graphs, the first two theorems are motivated by models on regular $D$-dimensional lattices where the interaction strength between two sites can be upper-bounded by a power law proportional to $d(i,j)^\alpha$, where $d(i,j)$ denotes the graph distance between lattice sites $i$ and $j$, and $\alpha\geq0$ is an exponent characterizing the spatial decay. For such regular lattices, Theorem 1 provides a Lieb-Robinson bound for interactions with $\alpha\geq D$, whereas Theorem 2 extends the applicability to $\alpha$ between zero and $D$, at the expense of having to work in suitably rescaled time. Both these bounds have a simple functional form, similar to that of Lieb and Robinson's original bound \eqref{e:LR_general}, but with an algebraic spatial decay instead of an exponential one. Theorem 3 is complementary in the sense that it provides a bound that is tighter than those of the first two theorems, but less explicit, requiring the evaluation of the exponential of an $N\times N$-matrix, where $N$ is the number of lattice sites (or graph vertices). 
Bosonic systems, for which the local Hilbert space dimension is infinite, are not covered by any of the above mentioned results. Our Theorem 4 fills this gap for open harmonic lattice models with long-range interactions by providing bounds on the norm of commutators between canonical coordinates. The tools required to prove all these bounds combine super-operator norms suitable for open quantum systems with techniques developed for Lieb-Robinson bounds for unitarily evolving quantum systems with long-ranged interactions. We provide in Section \ref{s:setting} an introduction to the relevant setting and tools, following closely the presentation of Refs.~\cite{Barthel_OQSLR_2012,KlieschGogolinEisert14}, before 
presenting the 
main results, as described above, in Section \ref{s:LRbounds}. 

\section{Setting}
\label{s:setting}

\subsection{The adjoint quantum master equation}

We are interested in open quantum systems whose evolution can be described by a differentiable evolution family of quantum channels \cite{Rivas_OQS_2012}. Given a Hilbert space $\mathcal{H}$, we denote the set of all bounded linear operators $A:\mathcal{H}\rightarrow \mathcal{H}$ as $\mathcal{B}(\mathcal{H})$. We are then interested in systems for which the state of the system $\rho(t)\in \mathcal{B}(\mathcal{H})$, at time $t \geq s$, is given by $\rho(t) = T(t,s)\rho(s)$, where $T(t_2,t_1):\mathcal{B}(\mathcal{H}) \rightarrow \mathcal{B}(\mathcal{H}) \in \mathcal{B}(\mathcal{B}(H))$ is a completely positive, trace preserving map (a quantum channel) for all $ t_1\leq t_2 \in \RR$, and $\{T(t,s)\}$ is an evolution family, i.e.
\begin{align}
T(t+r,s) = T(t+r,t)T(t,s),\qquad
T(s,s) = \mathds{1},
\end{align}
for all $s\leq t \leq t+r$. Assuming the evolution family to be differentiable, by differentiation we obtain
\begin{align}
\frac{d}{dt}T(t,s) = \tilde{\mathcal{L}}(t)T(t,s),\qquad
\frac{d}{ds}T(t,s) = -T(t,s)\tilde{\mathcal{L}}(s),
\end{align}
where
\begin{equation}
\tilde{\mathcal{L}}(t) = \lim_{\epsilon \rightarrow 0}\frac{T(t+\epsilon,t) - \mathds{1}}{\epsilon}
\end{equation}
is the generator of the evolution family. One can verify that a solution to these equations is provided by the time-ordered exponential of the generator,
\begin{equation}\label{propagators}
T(t,s) = \overleftarrow{T}\mathrm{exp}\Big(\int_s^{t}\tilde{\mathcal{L}}(t')dt'   \Big).
\end{equation}
Using all of the above, we find that the dynamics of the system's state $\rho$ satisfies the quantum master equation 
\begin{equation}\label{master}
\frac{d}{dt}\rho(t) = \tilde{\mathcal{L}}(t)\rho(t).
\end{equation}
The right-hand side of this equation can always be cast in the diagonal time-dependent Gorini-Kossakowski-Sudarshan-Lindblad (GKSL) form
\begin{align}
\tilde{\mathcal{L}}(t)\rho(t) &= -i[H(t),\rho(t)] + \sum_{v = 1}^{M} \gamma_v(t)\Big[\tilde{L}_v(t)\rho(t)\tilde{L}^{\dagger}_v(t) - \frac{1}{2}\Big( \tilde{L}_v^{\dagger}(t)\tilde{L}_v(t)\rho(t) + \rho(t)\tilde{L}_v^{\dagger}(t)\tilde{L}_v(t)\Big)\Big] \\
 &\equiv -i[H(t),\rho(t)] + \sum_{v = 1}^{M}\Big[ L_v(t)\rho(t)L^{\dagger}_v(t) - \frac{1}{2}\Big( L_v^{\dagger}(t)L_v(t)\rho(t) + \rho(t)L_v^{\dagger}(t)L_v(t)\Big)\Big], \label{e:GKSL}
\end{align}
where $H(t) \in \mathcal{B}(\mathcal{H})$ is a 
time-dependent Hamiltonian, $\{\tilde{L}_v(t)\in\mathcal{B}(\mathcal{H})\}_{v=1}^M$ with $M \leq \mathrm{dim}(\mathcal{H})^2 -1$ is a set of 
time-dependent Lindblad operators describing dissipation processes, and $\gamma_v(t)$ are the dissipation rates satisfying $\gamma_v(t) \geq 0$ for all $v$ and $t$, which is necessary to ensure that the propagators in Eq.~\eqref{propagators} are completely positive for all $s$ and $t$ \cite{Rivas_OQS_2012, Petruccione_OQS_2002}.

For the purpose of deriving a Lieb-Robinson bound we are interested in the time evolution of observables. To this end it is convenient to derive the quantum master equation in the Heisenberg picture (also known as the adjoint quantum master equation). Given the Hilbert-Schmidt inner product $\langle A,B\rangle_{\mathrm{HS}} = \mathrm{Tr}(A^{\dagger} B)$ on $\mathcal{B}(\mathcal{H})$, 
we define the adjoint of a super-operator $T$ as the map $T^{\dagger} \in \mathcal{B}(\mathcal{B}(\mathcal{H}))$ which satisfies $\langle T(A),B\rangle_{\mathrm{HS}} = \langle A, T^{\dagger}(B)\rangle_{\mathrm{HS}}$ for all $A,B \in \mathcal{B}(\mathcal{H})$. This definition implies that, if $\{T(t,s)\}$ is a differentiable evolution family, then $\{T^{\dagger}(t,s)\}$ is a \textit{backwards} differentiable evolution family, i.e. 
\begin{align}\label{back1}
T^{\dagger}(t+r,s) = T^{\dagger}(t,s)T^{\dagger}(t+r,t),\qquad
T^{\dagger}(t,t) = \mathds{1},
\end{align}
for all $s\leq t \leq t+r$. Given an observable $A \in \mathcal{B}(\mathcal{H})$ and $ s \leq t$, it is then natural to define $A(t):= A$ and consider the backward time-evolved observables $A(s) := T^{\dagger}(t,s)A$ such that
\begin{align}
\langle A\rangle_{s\rightarrow t} = \mathrm{Tr}\left[\rho(t)A\right] = \mathrm{Tr}\left[\left(T(t,s)\rho(s)\right)A\right] =\mathrm{Tr}\left[\rho(s)\left(T^{\dagger}(t,s)A\right)\right] := \mathrm{Tr}\left[\rho(s)A(s)\right].
\end{align}
In order to derive an equation of motion for $A(s)$ we use Eq.~\eqref{back1} to differentiate $T^{\dagger}(t,s)$, from which we obtain
\begin{align}
\frac{d}{dt}T^{\dagger}(t,s) = T^{\dagger}(t,s)\tilde{\mathcal{L}}^{\dagger}(t), \label{adeq1}\qquad
\frac{d}{ds}T^{\dagger}(t,s) = -\tilde{\mathcal{L}}^{\dagger}(s)T^{\dagger}(t,s),
\end{align}
where $\tilde{\mathcal{L}}^{\dagger}(t)$, the adjoint of $\tilde{\mathcal{L}}(t)$, is the generator of the backwards evolution family, given by 
\begin{equation}
\tilde{\mathcal{L}}^{\dagger}(t) = \lim_{\epsilon \rightarrow 0}\frac{T^{\dagger}(t+\epsilon,t) - \mathds{1}}{\epsilon}.
\end{equation}
In this case one can verify that 
\begin{equation}
T^{\dagger}(t,s) = \overrightarrow{T}\mathrm{exp}\Big(\int_s^{t}\tilde{\mathcal{L}}^{\dagger}(t')dt'   \Big),
\end{equation}
the \textit{backwards} time-ordered exponential of the adjoint generator $\tilde{\mathcal{L}}^{\dagger}(t)$, provides a solution to \eqref{adeq1} with initial condition specified in \eqref{back1}. Combining all of the above, we obtain that backward time-evolved observables in the Heisenberg picture satisfy the adjoint quantum master equation
\begin{equation}\label{HE1}
\frac{d}{ds}A(s) = -\tilde{\mathcal{L}}^{\dagger}(s)A(s),
\end{equation}
where
\begin{equation}\label{HE2}
\tilde{\mathcal{L}}^{\dagger}(s)A(s) = i[H(s),A(s)] + \sum_v \left[L^{\dagger}_v(s)A(s)L_v(s) - \frac{1}{2}\Big( L_v^{\dagger}(s)L_v(s)A(s) + A(s)L_v^{\dagger}(s)L_v(s)\Big)\right].
\end{equation}
As we will only be working in the Heisenberg picture\je{,} it is convenient to introduce the notation $\tau(s,t) := T^{\dagger}(t,s)$ and $\mathcal{L}(s) := \tilde{\mathcal{L}}^{\dagger}(s)$, which lets us summarize concisely as follows: Given an open quantum system described by a differentiable evolution family of quantum channels $\{T(t,s)\}$ (i.e. a system satisfying a time-dependent GKSL master equation), which is conventionally specified in terms of the generator $\tilde{\mathcal{L}}(t)$, then for any observable $A \in \mathcal{B}(\mathcal{H})$ and any initial 
state $\rho(s)$, we have that for any $s \leq t$ 
\begin{align}
\langle A\rangle_{s\rightarrow t} = \mathrm{Tr}(\rho(t)A) = \mathrm{Tr}(\rho(s)A(s)),
\end{align} 
where $\rho(t) = T(t,s)\rho(s)$, $A(s) =\tau(s,t)A$ 
and
\begin{equation}\label{adjointmaster}
\frac{d}{ds}A(s) = -\mathcal{L}(s)A(s).
\end{equation}

\subsection{Norms}
\label{norms}

Given $A \in \mathcal{B}(\mathcal{H})$ we define the Schatten $p$-norm as
\begin{equation}
\lVert A\rVert_p := \Big[\mathrm{Tr}\big[(A^{\dagger}A)^{p/2}   \big]   \Big]^{1/p}.
\end{equation}
We will utilize the Schatten 1-norm, or trace norm,
\begin{equation}
\lVert A\rVert_1 := \lVert A\rVert_\mathrm{tr} = \mathrm{Tr}\sqrt{A^{\dagger}A},
\end{equation}
as well as the $\infty$-norm, or operator norm,
\begin{equation}
\lVert A\rVert_\infty := \lim_{p\rightarrow\infty}\lVert A\rVert_p := \lVert A\rVert =  \sup_{x \in \mathcal{H}}\frac{\lVert A(x)\rVert}{\lVert x\rVert}.
\end{equation}
The trace norm is the physically most relevant norm for quantum states, while the operator norm is the physically relevant norm for observables. Given the $p$-norms we can now define the induced $p\rightarrow q$ super-operator norm via
\begin{equation}
\lVert T\rVert_{p \rightarrow q} := \sup_{A \in \mathcal{B}(\mathcal{H})}\frac{\lVert T(A)\rVert_p}{\lVert A\rVert_q}
\end{equation}
for all $T \in \mathcal{B}(\mathcal{B}(\mathcal{H}))$. Again, we will be interested in both the $\infty\rightarrow\infty$ norm (when working in the Heisenberg picture) as well as the $1\rightarrow 1$ norm (when working in the Schr\"{o}dinger picture). The following norm properties (amongst other generic properties of norms) are used for the proofs of Lieb-Robinson bounds.
\begin{enumerate}
\renewcommand{\labelenumi}{(\roman{enumi})}
\item $\lVert AB\rVert_p \leq \lVert A\rVert_p\lVert B\rVert_p$ for all  $A,B \in \mathcal{B}(\mathcal{H})$ (submultiplicativity of the $p$-norms \cite{watrous_QIT_2018}).
\item $\lVert UAV^{\dagger}\rVert_p = \lVert A\rVert_p$ for all $A\in \mathcal{B}(\mathcal{H})$ and for all unitary $U,V \in \mathcal{B}(\mathcal{H})$ (unitary invariance of the $p$-norms \cite{watrous_QIT_2018}).
\item $\lVert TQ\rVert_{p\rightarrow p} \leq \lVert T\rVert_{p\rightarrow p} \lVert Q\rVert_{p\rightarrow p}$ for all $T,Q \in \mathcal{B}(\mathcal{B}(\mathcal{H}))$ (submultiplicativity of the $p\rightarrow p$ norms \cite{watrous_QIT_2018}).
\item $\lVert T(A)\rVert_p \leq \lVert T\rVert_{p\rightarrow q}\lVert A\rVert_q$ for all $T \in \mathcal{B}(\mathcal{B}(\mathcal{H}))$ and for all $A\in \mathcal{B}(\mathcal{H})$ \cite{watrous_QIT_2018}.
\item $\lVert T\rVert_{1\rightarrow 1} = \lVert T^{\dagger}\rVert_{\infty\rightarrow\infty}$ for all $T \in \mathcal{B}(\mathcal{B}(\mathcal{H}))$ (``duality" of $1\rightarrow 1$ and $\infty\rightarrow\infty$ norms) \cite{Barthel_OQSLR_2012}.
\item $\lVert T\rVert_{1\rightarrow1} = 1$ for any quantum channel $T \in \mathcal{B}(\mathcal{B}(\mathcal{H}))$ \cite{Kliesch_DQCT_2011}.
\end{enumerate}

Some subtleties associated with the $p\rightarrow q$ induced super-operator norms are worth being mentioned. Firstly, the $1\rightarrow 1$ norm is not stable with respect to tensoring with the identity, i.e. there exist super-operators $T \in \mathcal{B}(\mathcal{B}(\mathcal{H}))$ such that
\begin{equation}
\lVert T\rVert_{1\rightarrow 1} \neq \lVert T\otimes \mathds{1}_{\mathcal{B}(\mathcal{H}_2)}\rVert_{1\rightarrow 1},
\end{equation}
 where $\mathds{1}_{\mathcal{B}(\mathcal{H}_2)}$ is the identity super-operator in $\mathcal{B}(\mathcal{B}(\mathcal{H}_2))$ for some Hilbert space $\mathcal{H}_2$ \cite{Watrous_2005, watrous_QIT_2018}. This is important as we will often be working with super-operators of the form $T = \tilde{T}\otimes\mathds{1}$, and one needs to ensure that one does not assume $\lVert T\rVert_{1\rightarrow 1} = \lVert \tilde{T}\rVert_{1\rightarrow 1} $. Similarly, there exist super-operators $T \in \mathcal{B}(\mathcal{B}(\mathcal{H}))$ and Hilbert spaces $\mathcal{H}_2$ and $\mathcal{H}_3$ such that
\begin{equation}
\lVert T\otimes \mathds{1}_{\mathcal{B}(\mathcal{H}_2)}\rVert_{1\rightarrow 1} \neq \lVert T\otimes \mathds{1}_{\mathcal{B}(\mathcal{H}_3)}\rVert_{1\rightarrow 1}.
\end{equation}
In particular, given an arbitrary $T \in \mathcal{B}(\mathcal{B}(\mathcal{H}))$ one can show \cite{Watrous_2005} that, if $\mathrm{dim}(\mathcal{H}_2) \geq \mathrm{dim}(\mathcal{H})$, then one always has
\begin{equation}
\lVert T\otimes \mathds{1}_{\mathcal{B}(\mathcal{H}_2)}\rVert_{1\rightarrow 1} = \lVert T\otimes \mathds{1}_{\mathcal{B}(\mathcal{H})}\rVert_{1\rightarrow 1}.
\end{equation}
On the other hand, if $\mathrm{dim}(\mathcal{H}_2)  < \mathrm{dim}(\mathcal{H})$ and $\mathrm{dim}(\mathcal{H}_2) < \mathrm{dim}(\mathcal{H}_3)$, then it may be that 
\begin{equation}
\lVert T\otimes \mathds{1}_{\mathcal{B}(\mathcal{H}_2)}\rVert_{1\rightarrow 1} \leq \lVert T\otimes \mathds{1}_{\mathcal{B}(\mathcal{H}_3)}\rVert_{1\rightarrow 1}.
\end{equation}
We will have to take these properties into account in Section \ref{locality}.

\subsection{Lattice systems and many-body Liouvillians}\label{locality}

We consider finite lattices $\Lambda$, equipped with some metric $d$.
A finite Hilbert space $\mathcal{H}_x$ is associated to each $x \in \Lambda$, and we define $\mathcal{H}_X = \bigotimes_{x \in X}\mathcal{H}_x$ for all subsets $X \subset \Lambda$, and $\mathcal{H}:=\mathcal{H}_\Lambda$. 
Later we will extend our considerations to harmonic systems, for which 
each site is equipped with a harmonic bosonic mode, however the details of this setting are postponed till Section \ref{ss:harmonic}.
Given $A \in \mathcal{B}(\mathcal{H})$ we define the support of $A$, denoted $\mathrm{supp}(A)$, as the smallest subset $X\subset \Lambda$ for which there exists a non-trivial $A_X \in \mathcal{B}(\mathcal{H}_X)$ such that $A = A_X\otimes \mathds{1}_{\Lambda/X}$. For any $X \subset \Lambda$ we then define
\begin{equation}
\mathcal{B}_X(\mathcal{H}) := \{A \in \mathcal{B}(\mathcal{H})|
\mathrm{supp}(A)\subseteq X\}
\end{equation}
as the space of all bounded linear operators on $\mathcal{H}$ with support contained in $X$. We define the support of a super-operator $\mathcal{L}\in\mathcal{B}(\mathcal{B}(\mathcal{H}))$ as
\begin{equation}\label{e:support_1}
\mathrm{supp}(\mathcal{L}) := \bigcap \{X \subset \Lambda \mid \mathcal{B}_{\Lambda/X}(\mathcal{H}) \subseteq \mathrm{ker}(\mathcal{L})\}.
\end{equation}
We further define
\begin{equation}\label{e:support_2}
\mathds{L}'_X = \{\mathcal{L} \in \mathcal{B}(\mathcal{B}(\mathcal{H})) \mid \mathrm{supp}(\mathcal{L}) = X\}
\end{equation}
as the set of super-operators whose support is $X$,
and
\begin{equation}
\mathds{L}_X = \{\mathds{L}'_Y \mid Y \subseteq X\}
\end{equation}
as the set of super-operators whose support is a subset of $X$.

We are interested in open many-body quantum systems whose dynamics satisfies Eq.~\eqref{adjointmaster} for some adjoint generator $\mathcal{L}(s)$ that can be written as a sum of terms,
\begin{equation}\label{local}
\mathcal{L}(s) = \sum_{Z \subset \Lambda} \mathcal{L}_Z(s),
\end{equation}
where $\mathcal{L}_Z(s) \in \mathds{L}'_Z$ is the generator of a backwards differentiable evolution family in its own right. Typically, three categories of interactions are distinguished.
\begin{enumerate}
\renewcommand{\labelenumi}{(\alph{enumi})}
\item Short-range interactions \cite{Barthel_OQSLR_2012}: $\mathcal{L}$ is the sum of terms with finite norm bound 
\begin{equation}\label{e:normbound}
l := \sup_{s,Z\subset\Lambda}\lVert \mathcal{L}_Z(s)\rVert_{\infty\rightarrow\infty}, 
\end{equation}
finite maximum range
\begin{equation}
a := \sup_{Z:\mathcal{L}_Z\neq 0}\mathrm{diam}(Z),
\end{equation}
and finite maximum number of nearest neighbours
\begin{equation}
\mathcal{Z} := \max_{Z:\mathcal{L}_Z\neq 0}|\{Z'\subset\Lambda|\mathcal{L}_{Z'}\neq 0, Z'\cap Z \neq \emptyset \}|,
\end{equation}
where $\mathrm{diam}(Z) = \max_{x,y \in Z}d(x,y)$.
\item Exponentially-decaying interactions \cite{Nachtergaele_2011,HastingsKoma06}: There exist  positive constants $\lambda_0$ and $\mu$ such that for all $x,y \in \Lambda$ and for all $t \in \RR$,
\begin{equation}\label{e:expdecay}
\sum_{X \ni x,y} \sup_{s \in [0,t]}\lVert \mathcal{L}_X(s)\rVert_{\infty\rightarrow\infty} \leq \lambda_0 e^{-(\mu d(x,y))}.
\end{equation}
\item Power law-decaying interactions \cite{Nachtergaele_2011, HastingsKoma06}: There exist positive constants $\lambda_0$ and $\eta$ such that for all $x,y \in \Lambda$ and for all $t \in \RR$,
\begin{equation}\label{power}
\sum_{X \ni x,y} \sup_{s \in [0,t]} \lVert \mathcal{L}_X(s)\rVert_{\infty\rightarrow\infty} \leq \frac{\lambda_0}{[1 + d(x,y)]^{\eta}}.
\end{equation}
\end{enumerate}

Lieb-Robinson bounds for open quantum systems with adjoint generators of type (a) have been proved in Refs.~\cite{Poulin10,Barthel_OQSLR_2012}, and similar, but more general, results for adjoint generators of type (a) and (b) have been proven in Ref.~\cite{Nachtergaele_2011}. Here we are concerned with adjoint generators of type (c). As in the proofs of Lieb-Robinson bounds for closed (unitary) quantum systems with power law-decaying interactions, we need additional assumptions on the lattice $\Lambda$.
\newline

\noindent\textbf{Assumption 1:} The lattice $\Lambda$, equipped with metric $d$, satisfies
\begin{equation}\label{consequence}
\sum_{z \in \Lambda} \frac{1}{[1 +d(x,z)]^{\eta}}
\frac{1}{[1 +d(z,y)]^{\eta}} \leq \frac{p_0}{[1 +d(x,y)]^{\eta}}
\end{equation}
\textit{for some $p_0>0$.}

As shown in Ref.~\cite{HastingsKoma06}, if for a given positive $\eta$ we have that

\begin{equation}\label{extensivity}
\sup_{x \in \Lambda} \sum_{y \in \Lambda}\frac{1}{[1 +d(x,y)]^{\eta}} < \infty,
\end{equation}
then Eq.~\eqref{consequence} will hold for the same value of $\eta$. Eq.~\eqref{extensivity} therefore provides a simpler sufficient criterion for ensuring that Assumption 1 is satisfied. For a conventional metric on a $D$-dimensional regular lattice, Eq.~\eqref{extensivity} is violated for $\eta < D$ \cite{StorchvandenWormKastner15,MetivierBachelardKastner14}, and therefore Assumption 1 cannot be used to prove Lieb-Robinson bounds in that case. For power law-decaying interactions with $\eta < D$, we therefore follow Refs.\ \cite{StorchvandenWormKastner15,MetivierBachelardKastner14} and define an alternative assumption, which holds on regular $D$-dimensional lattices for all $\eta > 0$, and allows one, as shown in Section \ref{extensions}, to prove a Lieb-Robinson bound with a rescaled notion of time.\newline 

\noindent\textbf{Assumption 2:} \textit{The lattice $\Lambda$, equipped with metric $d$, satisfies}
\begin{equation}\label{consequence2}
\mathcal{N}_\Lambda\sum_{z \in \Lambda} \frac{1}{[1 +d(x,z)]^{\eta}}
\frac{1}{[1 +d(z,y)]^{\eta}} \leq \frac{p_1}{[1 +d(x,y)]^{\eta}},
\end{equation}
\textit{for some finite $p_1 > 0$ and for all  $x,y \in \Lambda$, where}
\begin{equation}\label{rescaling_factor}
\mathcal{N}_\Lambda = 1/ \sup_{x \in \Lambda}\sum_{y \in \Lambda/\{x\}}\frac{1}{[1 + d(x,y)]^\eta}.
\end{equation}
\newline

\noindent Finally, note that conditions (a)--(c) have been specified in terms of the $\infty \rightarrow \infty$ norm for adjoint generators. However, as discussed in Section \ref{norms}, super-operator norms are not stable with respect to tensoring the identity, which can create problems if one wishes to obtain results for dynamics defined not for a fixed finite lattice as we do here, but rather for a family of lattices---such as if one wishes to extend these results in a consistent manner to infinite lattice systems, as per the methods in Ref.~\cite{Nachtergaele_2011} for systems with exponentially decaying interactions. In particular, as a consequence of the way that adjoint generators with restricted support have been defined in Eqs.~\eqref{e:support_1} and \eqref{e:support_2}, we see that if $\mathcal{L}_Z \in \mathds{L}'_Z$, then there exists some $\hat{\mathcal{L}}_Z \in \mathcal{B}(\mathcal{B}(\mathcal{H}_Z))$ such that
\begin{equation}
\mathcal{L}_Z = \hat{\mathcal{L}}_Z\otimes\mathds{1}_{\mathcal{B}(\mathcal{H}_{\Lambda/Z})}.
\end{equation}
Therefore in general $\lVert \mathcal{L}_Z\rVert_{\infty\rightarrow\infty}$, and hence the constants appearing in conditions (a)--(c) depend on $\Lambda$. While we will not make use of this here, it is interesting to note that this dependence can be removed by defining the stabilized diamond norm
\begin{equation}
\lVert T\rVert_{\lozenge} = \lVert T\otimes\mathds{1}_{\mathcal{B}(\mathcal{H})}\rVert \quad \forall T \in \mathcal{B}(\mathcal{B}(\mathcal{H}))
\end{equation}
and by replacing $\lVert \mathcal{L}_X(s)\rVert_{\infty\rightarrow\infty}$ in \eqref{e:normbound}, \eqref{e:expdecay}, and \eqref{power} with the  completely bounded norm
\begin{equation}
\lVert \mathcal{L}_X(s)\rVert_\mathrm{cb} = \lVert \hat{\mathcal{L}}_X(s)\otimes\mathds{1}_{\mathcal{B}(\mathcal{H}_{\Lambda/X})}\rVert_{\mathrm{cb}} {}:= \lVert \hat{\mathcal{L}}_X(s)\rVert_{\lozenge},
\end{equation}
which no longer depends on $\Lambda$.

\section{Lieb-Robinson bounds}
\label{s:LRbounds}

\subsection{Lieb-Robinson bound for long-range interactions with $\eta>D$}\label{extensions0}

Lieb-Robinson bounds for open quantum systems have first been proven for the case of short-range interactions in Refs.~\cite{Poulin10,Barthel_OQSLR_2012}, via a natural generalization of the methods in Refs.~\cite{NachtergaeleSims06, NachtergaeleOgataSims06, HastingsKoma06}, by making use of the adjoint quantum master equation and the relevant super-operator norms. As noted in Ref.~\cite{Barthel_OQSLR_2012}, no formal obstacle prevents one from applying the methods of Refs.\ \cite{Poulin10, Barthel_OQSLR_2012}, in conjunction with the methods of Refs.~\cite{NachtergaeleSims06, NachtergaeleOgataSims06, HastingsKoma06}, to obtain bounds for systems with long-range interactions. Such an extension for open quantum systems with exponentially decaying interactions has been reported in Ref.~\cite{Nachtergaele_2011}. Below we provide an extension to open quantum systems with power law-decaying interactions, using techniques developed in Ref.~\cite{HastingsKoma06} in the context of closed systems.\newline

\noindent\textbf{Theorem 1 (Lieb-Robinson bound):} \textit{Given a finite lattice $\Lambda$ equipped with a metric $d$, and an open quantum system described by a differentiable evolution family of quantum channels whose adjoint generator can be written as in \eqref{local}, then if there exist positive constants $\lambda_0$ and $\eta$ such that \eqref{power} is satisfied for all $x,y \in \Lambda$ and for all $t \in \RR$, and if Assumption 1 is satisfied for the same value of $\eta$, then for any $K_X \in \mathds{L}_X$, $O_Y \in \mathcal{B}_{Y}(\mathcal{H})$ with $X\cap Y = \emptyset$, and $0 \leq r \leq t \in \RR$, we have that}
\begin{equation}\label{statement}
\lVert K_X\tau(r,t)O_Y\rVert \leq \frac{C\big(e^{v(t-r)} - 1  \big)}{[1+d(X,Y)]^{\eta}},
\end{equation}
\textit{where} 
\begin{equation}
C = \lVert K_X\rVert_{\infty\rightarrow\infty} \Vert O_Y\rVert|X||Y|p_0^{-1},\qquad
v = \lambda_0p_0,\qquad
d(X,Y) = \min_{x \in X, y \in Y}d(X,Y).
\end{equation}
Note that a more conventional form of the above bound is obtained by 
taking $K_X$ such that, for all $A \in\mathcal{B}(\mathcal{H})$,
\begin{equation}\label{eq:lie_operator}
K_X(A) = [A,O_X]
\end{equation}
for some $O_X \in \mathcal{B}_X(\mathcal{H})$. In this case we can replace $\lVert K_X\rVert_{\infty\rightarrow\infty}$ with $2\lVert O_X\rVert$. Stated more explicitly, under the same conditions as Theorem 1, but with $K_X$ as in eq. \eqref{eq:lie_operator}, for $O_X \in \mathcal{B}_{X}(\mathcal{H})$ one obtains
\begin{equation}
    \lVert [\tau(r,t)O_Y, O_X]\rVert \leq \frac{C\big(e^{v(t-r)} - 1  \big)}{[1+d(X,Y)]^{\eta}},
\end{equation}
with
\begin{equation}
C = 2\Vert O_X\Vert \Vert O_Y\rVert|X||Y|p_0^{-1}.
\end{equation}
\newline

\proof We define the quantity
	\begin{equation}\label{e:Gr}
		G(r) := K_X\tau(r,t)O_Y,
	\end{equation}	
whose norm we would like to bound. Similar to Refs.~\cite{NachtergaeleSims06,NachtergaeleOgataSims06, HastingsKoma06, Nachtergaele_2011, Poulin10, Barthel_OQSLR_2012}, the first step is to ``differentiate and integrate'' both sides of Eq.~\eqref{e:Gr}. Following the approach used in Ref.\ \cite{Barthel_OQSLR_2012} for open quantum systems with short-range interactions, we note that $G(t) = K_XO_Y$ and
	\begin{align}
		\frac{\partial}{\partial r}G(r) = - K_X\mathcal{L}(r)\tau(r,t)O_Y, = -\mathcal{L}_{\Lambda/X}(r)G(r) - K_X\mathcal{L}_{\bar{X}}(r)\tau(r,t),
	\end{align}
where we have utilized \eqref{back1} and \eqref{adeq1}. In a slight abuse of notation we have denoted $\mathcal{L}_{\Lambda/X} := \sum_{Z \subseteq \Lambda/X}\mathcal{L}_Z$ and $\mathcal{L}_{\bar{X}} := \sum_{Z \in \bar{X}}\mathcal{L}_Z$ with $\bar{X} := \{Z\in \Lambda \mid Z\cap X \neq \emptyset\}$, which implies that $[K_X,\mathcal{L}_{\Lambda/X}] = 0$ for all $K_X \in \mathds{L}_X$. A solution for $G(r)$ under these conditions is then given by 
	\begin{equation}
		G(r) = \tau_{\Lambda/X}(r,t)G(t) + \int_{r}^t ds\,\tau_{\Lambda/X}(r,s)K_X\mathcal{L}_{\bar{X}}(s)\tau(s,t)O_Y.
	\end{equation}
Taking the norm of $G(r)$ and utilizing the norm properties detailed in Section \ref{norms} allows us to obtain
	\begin{equation}\label{GNorm}
		\lVert  G(r) \rVert \leq \lVert  G(t) \rVert + \lVert K_X\rVert \sum_{Z \in \bar{X}} \int_r^t ds \lVert \mathcal{L}_Z(s)\tau(s,t)O_Y\rVert,
	\end{equation}
where we are using the short-hand notation $\lVert \bullet \rVert$ to denote $\lVert \bullet\rVert_{\infty\rightarrow\infty}$ for all norms of super-operators. From here we proceed by defining the related quantity
	\begin{equation}\label{e:commutatorfraction}
		C_X(r) := \sup_{K \in \mathds{L}_X} \frac{\lVert K\tau(r,t)O_Y\rVert}{\lVert K\rVert},
	\end{equation}
which we will be able to bound. Dividing Eq.~\eqref{GNorm} by $\lVert K_X\rVert$ and taking the supremum gives us
	\begin{equation}\label{iterationeq}
		C_X(r) \leq C_X(t) + \sum_{Z \in \bar{X}} \sup_{s\in [r,t]}\lVert \mathcal{L}_{Z}(s)\rVert\int_r^tdsC_Z(s).	
	\end{equation}
Utilizing properties of the norm, we note that
	\begin{equation}
		C_X(t) \leq \delta(X,Y)\lVert O_Y\rVert,
	\end{equation}
where $\delta(X,Y) := 0$ if $X\cap Y = \emptyset$ or 1 otherwise. Assuming $X\cap Y = \emptyset$ and substituting \eqref{iterationeq} into itself and iterating yields the expression
	\begin{equation}\label{infsum}
		C_X(r) \leq \lVert O_Y\rVert\sum_{n = 1}^{\infty} \frac{(t-r)^n}{n!}c_n,
	\end{equation}
where
	\begin{equation}\label{e:powerseriesterms}
		c_n := \sum_{Z_1\in \bar{Z}_0}\sum_{Z_2\in \bar{Z}_1}\ldots\sum_{Z_n\in \bar{Z}_{n-1}}\delta(Z_n,Y)\prod_{i = 1}^n \sup_{s\in [r,t]} \lVert \mathcal{L}_{Z_i}(s)\rVert
	\end{equation}
with $Z_0 := X$.
To further bound the right-hand side of \eqref{infsum}, we derive upper bounds on the terms $c_n$. To achieve this, we adapt a method that was introduced in Ref.~\cite{HastingsKoma06} for proving Lieb-Robinson bounds for closed quantum systems with power law-decaying interactions. Similarly to
Ref.\ \cite{HastingsKoma06}, we note that
\begin{subequations}
	\begin{align}
		c_1 &= \sum_{Z_1\in\bar{Z}_0}\delta(Z_1,Y)\sup_{s\in [r,t]} \lVert \mathcal{L}_{Z_1}(s)\rVert \\
		&\leq \sum_{x\in X}\sum_{y \in Y}\sum_{Z \ni x,y} \sup_{s\in [0,t]} \lVert \mathcal{L}_{Z_1}(s)\rVert \\
		&\leq \sum_{x\in X}\sum_{y \in Y}\frac{\lambda_0}{[1+\mathrm{dist}(x,y)]^{\eta}}\\
		&\leq \frac{\lambda_0|X||Y|}{[1+\mathrm{dist}(X,Y)]^{\eta}},
	\end{align}
\end{subequations}
where we have used the power-law assumption \eqref{power}. 
By additionally using Assumption 1, $c_2$ can be bounded in a similar fashion,
\begin{subequations}
	\begin{align}
		c_2 &= \sum_{Z_1\in\bar{Z}_0}\sum_{Z_2\in\bar{Z_1}_0}\delta(Z_2,Y)\sup_{s\in [r,t]} \lVert \mathcal{L}_{Z_1}(s)\rVert\sup_{s\in [r,t]} \lVert \mathcal{L}_{Z_2}(s)\rVert \\
		&\leq \sum_{Z_1\in\bar{Z}_0}\sum_{Z_2\in\bar{Y}_0}\sup_{s\in [r,t]} \lVert \mathcal{L}_{Z_1}(s)\rVert\sup_{s\in [r,t]} \lVert \mathcal{L}_{Z_2}(s)\rVert \\
		&\leq \sum_{x\in X}\sum_{y \in Y}\sum_{z_{1,2} \in \Lambda}\sum_{Z_1 \ni z_{1,2},x}\sum_{Z_2 \ni z_{1,2},y} \sup_{s\in [0,t]} \lVert \mathcal{L}_{Z_1}(s)\rVert\sup_{s\in [0,t]} \lVert \mathcal{L}_{Z_2}(s)\rVert \\
		&\leq \sum_{x\in X}\sum_{y \in Y}\sum_{z_{1,2} \in \Lambda}\frac{\lambda_0}{[1+\mathrm{dist}(x,z_{1,2})]^{\eta}}\frac{\lambda_0}{[1+\mathrm{dist}(z_{1,2},y)]^{\eta}}\\
		&\leq \sum_{x\in X}\sum_{y \in Y} \frac{\lambda_0^2p_0}{[1+\mathrm{dist}(x,y)]^{\eta}}\\
		&\leq \frac{p_0\lambda_0^2|X||Y|}{[1+\mathrm{dist}(X,Y)]^{\eta}}.
	\end{align}
\end{subequations}
Proceeding in this manner, we find that 
	\begin{equation}\label{e:cn}
		c_n \leq \frac{p_0^{n-1}\lambda_0^n|X||Y|}{[1+\mathrm{dist}(X,Y)]^{\eta}}
	\end{equation}
for all $n$. Substituting \eqref{e:cn} into \eqref{infsum} gives
	\begin{equation}
		C_X(r) \leq \frac{\lVert O_Y\rVert|X||Y|}{p_0[1+\mathrm{dist}(X,Y)]^{\eta}}\left(e^{p_0\lambda_0(t-r)} - 1\right),
	\end{equation}
from which the theorem follows via the definition of $C_X(r)$. \proofend

\subsection{Lieb-Robinson bounds for arbitrarily long-ranged interactions}\label{extensions}

In the proof of Theorem 1, Assumption 1 has been crucial for bounding the coefficients $c_n$. However, for long-ranged interactions with $\eta < D$ and conventional metrics on the lattice,
Assumption 1 no longer holds, as discussed in Section \ref{locality}. Despite this, as shown in Ref.\ \cite{StorchvandenWormKastner15} for closed quantum systems, a Lieb-Robinson bound in rescaled time can be obtained by utilizing Assumption 2 instead of 1. Below we provide a similar result for open quantum systems.\newline

\noindent\textbf{Theorem 2 (Lieb-Robinson bound for arbitrarily long-ranged interactions):}\textit{ Given a finite lattice $\Lambda$ equipped with a metric $d$, and an open quantum system described by a differentiable evolution family of quantum channels whose adjoint generator can be written as in \eqref{local}, then, if there exist positive constants $\lambda_0$ and $\eta$ such that \eqref{power} is satisfied for all $t \in \RR$ and if Assumption 2 is satisfied for $\eta$, then for any $K_X \in \mathds{L}_X$, $O_Y \in \mathcal{B}_{Y}(\mathcal{H})$ with $X\cap Y = \emptyset$ and $0 \leq r \leq t \in \RR$ we have that}
\begin{equation}\label{statement2}
\lVert K_X\tau(r,t)O_Y\rVert \leq \frac{C_1\big(e^{[v_1(t-r)/\mathcal{N}_\Lambda}] - 1  \big)}{[1+d(X,Y)]^{\alpha}},
\end{equation}
\textit{where} 
\begin{align}
C_1 = \lVert K_X\rVert_{\infty\rightarrow\infty} \lVert O_Y\rVert|X||Y|\mathcal{N}_{\Lambda}p_1^{-1},\qquad
v_1 = \lambda_0p_1.
\end{align}

\proof Up until Eq.~\eqref{infsum}, the proof of Theorem 2 proceeds exactly as the proof of Theorem 1. At this point, as per \cite{StorchvandenWormKastner15}, by comparing expressions \eqref{consequence} and \eqref{consequence2} in Assumptions 1 and 2 respectively, we note that the rest of the proof proceeds identically, provided one replaces $p_0$ with $p_1/\mathcal{N}_{\Lambda}$, which allows for the use of Assumption 2 in place of Assumption 1. \proofend\newline

\noindent It is interesting to note that the bound on the right hand side of 
Eq.~\eqref{statement2} depends implicitly on the system size through the inclusion of the rescaling factor $\mathcal{N}_{\Lambda}$. However, from the definition of $\mathcal{N}_{\Lambda}$ in Eq.~\eqref{rescaling_factor} one can see that $\mathcal{N}_{\Lambda} \leq 2^{\eta}$, and it decreases with increasing system size. If desired it is therefore  possible to remove this system-size dependence by replacing $\mathcal{N}_{\Lambda}$ with $2^{\eta}$.

Another type of Lieb-Robinson bound, which has the form of a matrix exponential, was also put forward in Ref.~\cite{StorchvandenWormKastner15} in the context 
of closed quantum systems. It has the advantage of being tighter than other bounds, at the expense of having to calculate a certain matrix exponential. In the following theorem we generalize this matrix exponential bound to open quantum systems.\newline

\noindent\textbf{Theorem 3 (Matrix exponential bound for pair interactions):} \textit{Consider a finite lattice $\Lambda$ and an open quantum system described by a differentiable evolution family of quantum channels, whose adjoint generator can be written as the sum of symmetric pairwise terms, i.e.}
\begin{equation}\label{e:pairwise}
\mathcal{L}(s) = \frac{1}{2}\sum_{k \neq l}\mathcal{L}_{k,l}(s) = \sum_{k < l}\mathcal{L}_{k,l}(s) ,
\end{equation}
\textit{where $\mathrm{supp}(\mathcal{L}_{k,l}(s)) = \{k,l\}$. Then for $X = \{i\}$ and $Y = \{j\}$ with $i\neq j$ we have that for any single-site operators $K_X \in \mathds{L}_X$, $O_Y \in \mathcal{B}_{Y}(\mathcal{H})$ and any $0 \leq r \leq t \in \RR$,}
\begin{equation}\label{statement3}
\lVert K_X\tau(r,t)O_Y\rVert \leq \lVert K_X\rVert_{\infty \rightarrow \infty } \lVert O_Y\rVert  \left[ \exp(\kappa J (t-r))\right]_{i,j},
\end{equation}
where
\begin{equation}
J_{k,l} :=
  \begin{cases} 
   \sup_{s \in [r,t]} \lVert \mathcal{L}_{k,l}(s)\rVert_{\infty \rightarrow \infty} & \text{if  $k\neq l$},\\
   1   & \text{if $k=l$},
  \end{cases}
\end{equation}
and 
\begin{equation}
\kappa =  \sup_{n \in \Lambda} \sum_{k\neq n} J_{n,k}.
\end{equation}

\begin{figure}
  \centering
    \includegraphics[width=0.5\textwidth]{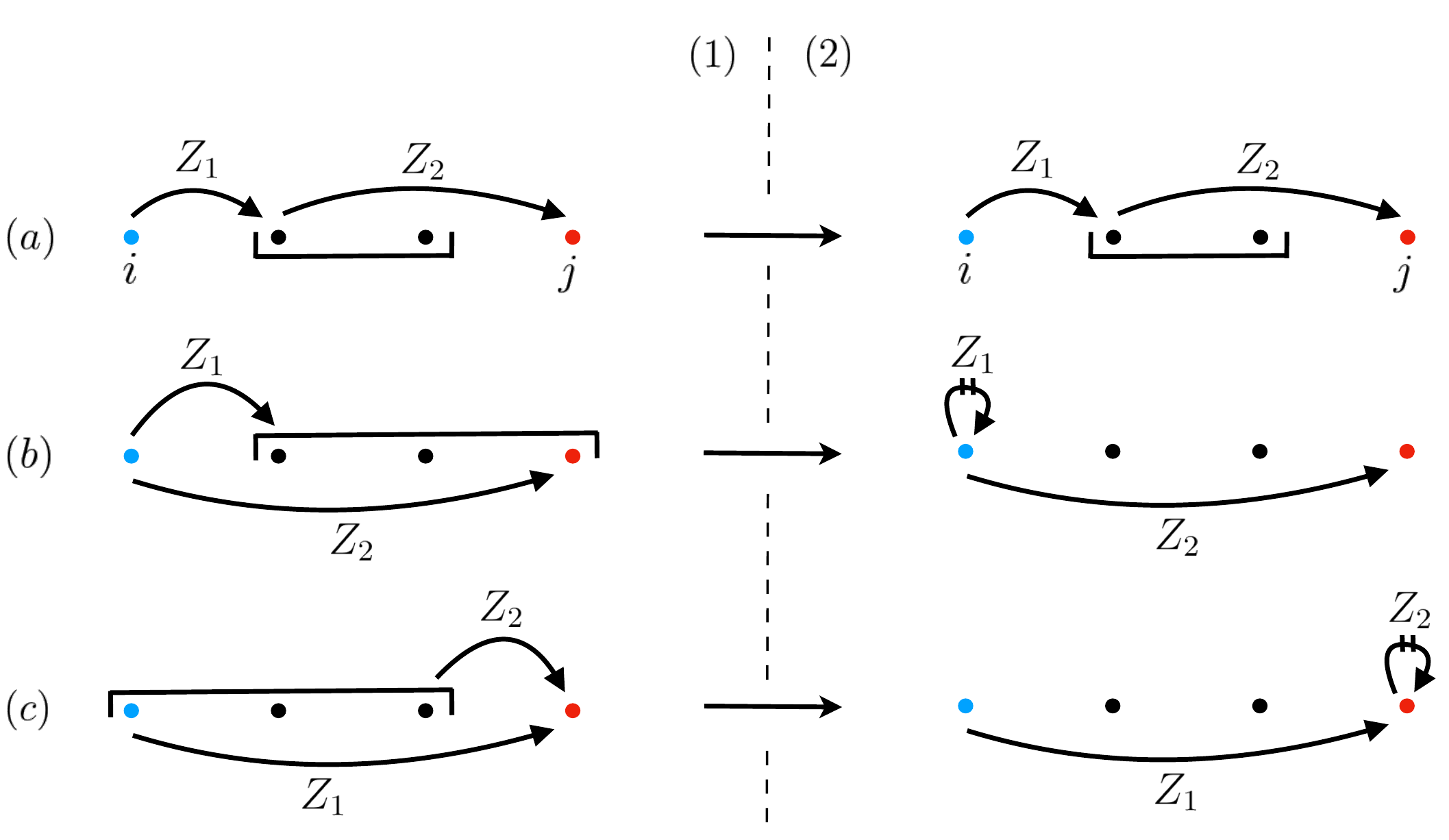}
    \caption{(1a)--(1c) Under the conditions of Theorem 3, the sum over intersecting subsets in the first line of \eqref{e:n2originalpaths} can be rewritten as the sum over three distinct subsets of all directed two-edge graphs admitting a path connecting sites $i$ and $j$, but in which no loops are present and in which the first edge is constrained to originate from site $i$ and the second edge is constrained to end on site $j$. (2a)--(2c) For each subset of graphs which contains ``free'' edges---i.e.\ edges which are not part of the path connecting sites $i$ and $j$---we can bound the sum over this subset of graphs by a single graph containing a loop with weight $\kappa$ on the common site on which the free edges begin or end.}\label{f:graph_paths}
\end{figure}

\proof Up until Eq.~\eqref{infsum} the proof of Theorem 3 proceeds as per the proof of Theorem 1. Then, as in Refs.~\cite{StorchvandenWormKastner15,GongFossFeigMichalakisGorshkov14}, for a pairwise adjoint generator in the form of Eq.~\eqref{e:pairwise} and for $X = \{i\}$ and $Y = \{j\}$ with $i\neq j$, one can show that the coefficients $c_n$ are upper bounded as
\begin{equation}\label{e:powerseriesbound}
c_n \leq \kappa^n [J^n]_{i,j}.
\end{equation}
For $n=1$ the bound in Eq.~\eqref{e:powerseriesbound} follows straightforwardly, and so in order to prove these bounds we start by looking at the case of $n=2$. In particular, as illustrated in the left-hand panel of Fig.~\ref{f:graph_paths} and shown in Eq.~\eqref{e:n2originalpaths}, for this case Eq.~\eqref{e:powerseriesterms} can naturally be written as the sum of three distinct contributions, each of which describes the sum over a subset of all the directed two-edge graphs admitting a path connecting sites $i$ and $j$, in which no loops are allowed (reflecting the fact that there are no on-site interaction terms in the adjoint generator) and in which the first edge is constrained to originate from site $i$ and the second edge is constrained to end on site $j$. Specifically, the first contribution describes the sum over the subset of graphs in which the second edge begins on the endpoint of the first edge, while the second and third contributions describes the sum over the subsets in which the second and first edges, respectively, connect sites $i$ and $j$ directly,
\begin{equation}
c_2 = \sum_{Z_1 \in \bar{\{i\}}}\sum_{Z_2 \in \bar{Z_1}}\delta(Z_2,\{j\})\sup_{s\in [r,t]} \lVert \mathcal{L}_{Z_1}(s)\rVert\sup_{s\in [r,t]} \lVert \mathcal{L}_{Z_2}(s)\rVert = \sum_{k \neq i,j} J_{i,k}J_{k,j} + \sum_{k\neq i} J_{i,k}J_{i,j} + \sum_{k\neq j}J_{i,j}J_{k,j}.\label{e:n2originalpaths}
\end{equation}
At this stage, as illustrated in the right-hand panel of Fig.~\ref{f:graph_paths} and shown explicitly below, for any subset of graphs defined by a common path connecting sites $i$ and $j$, but with edges beginning (or ending) on a particular site which are not part of this path, we can bound the sum over all such graphs by collecting the contribution over all edges originating from or ending on this site into a loop on this site with weight $\kappa$. As shown below in Eq.~\eqref{e:refinedpaths}, this procedure allows us to bound \eqref{e:n2originalpaths} as the sum over all directed two-edge graphs admitting a path connecting sites $i$ and $j$, in which loops with weight $\kappa$ are now allowed, but in which the second edge is constrained to originate on the end of the first edge. Using that we have defined $J_{k,k} := 1$, we see that
\begin{equation}\label{e:refinedpaths}
c_2 \leq \sum_{k \neq i,j} J_{i,k}J_{k,j} +  (\kappa J_{i,i})J_{i,j} + J_{i,j}(\kappa J_{j,j}) \leq \kappa^2 \sum_k J_{i,k}J_{k,j} = \kappa^2[J^2]_{i,j},
\end{equation}
which is the case $n=2$ of the bound \eqref{e:powerseriesbound} we want to show. For $n=3$ we can proceed in an analogous manner. We begin by rewriting the sum over intersecting subsets given by \eqref{e:powerseriesterms} as a sum over all the directed \textit{three}-edge graphs admitting a path connecting sites $i$ and $j$, in which no loops are allowed and in which the first edge is constrained to originate from site $i$ and the third edge is constrained to end on site $j$, and then continue by using the same trick as for the case of $n=2$ to bound this sum by the sum over all directed three-edge graphs admitting a path connecting sites $i$ and $j$, in which loops with weight $\kappa$ are allowed, but in which edge $m+1$ is constrained to begin at the endpoint of edge $m$ for $m \in \{2,3\}$. Explicitly, we find 
\begin{subequations}
\begin{align}
c_3 &= \sum_{Z_1 \in \bar{\{i\}}}\sum_{Z_2 \in \bar{Z_1}}\sum_{Z_3 \in \bar{Z_2}}\delta(Z_3,\{j\}) \prod_{k = 1}^3 \sup_{s\in [r,t]}\lVert \mathcal{L}_{Z_k}(s)\rVert \\
&=\sum_{k \neq i}\sum_{l \neq k,j}J_{i,k}J_{k,l}J_{l,j} 
 + \sum_{k \neq i}\sum_{l \neq i}J_{i,k}J_{i,l}J_{i,j} + \sum_{k \neq i}\sum_{l \neq j}J_{i,k}J_{i,j}J_{l,j} + \sum_{k \neq j}\sum_{l \neq j}J_{i,j}J_{k,j}J_{l,j} \nonumber \\
&  \qquad + \sum_{k \neq i}\sum_{l \neq i,j}J_{i,k}J_{i,l}J_{l,j} + \sum_{k \neq i, j}\sum_{l \neq k}J_{i,k}J_{l,k}J_{k,j} + \sum_{k \neq i, j}\sum_{l \neq k}J_{i,k}J_{k,j}J_{l,j} \\
& \leq \sum_{k \neq i}\sum_{l \neq k,j}J_{i,k}J_{k,l}J_{l,j}  + \kappa^2 \Big(J_{i,i}J_{i,i}J_{i,j} +  J_{i,i}J_{i,j}J_{j,j} +  J_{i,j}J_{j,j}J_{j,j} \Big)\nonumber \\
&  \qquad +\kappa\Big(\sum_{k\neq i,j}J_{i,i}J_{i,k}J_{k,j} + \sum_{k\neq i,j}J_{i,k}J_{k,k}J_{k,j} + \sum_{k\neq i,j}J_{i,k}J_{k,j}J_{j,j}\Big) \\
& \leq \kappa^3 \sum_{k,l}J_{i,k}J_{k,l}J_{l,j}  = \kappa^3 [J^3]_{i,j}.
\end{align}
\end{subequations}
An analogous treatment for larger $n$ yields the bound \eqref{e:powerseriesbound}. From this bound, together with Eqs.~\eqref{e:commutatorfraction} and \eqref{infsum}, Theorem 3 then follows. \proofend

While the bounds here are stated in terms of spin systems, by virtue of the
arguments of Ref.\ \cite{HastingsKoma06}, they equally apply, for observables that respect the superselection rule of the parity of fermion number, to fermionic lattice systems in which each site is associated not with a spin but with a fermionic degree of freedom. This is not entirely obvious, since a naive mapping of fermions to spins by means of a Jordan-Wigner transformation leads to non-local spin operators.

\subsection{Locality bounds for harmonic open quantum systems} \label{ss:harmonic}

In this section, we show that bounds similar to those of the previous two sections hold true also for harmonic lattice models, described by master equations in the GKSL form \eqref{e:GKSL} with Hamiltonians and Lindblad operators that are bilinear and linear respectively in bosonic creation and annihilation operators. These bounds generalize the results of Ref.~\cite{HarmonicLR}, which are valid for unitarily evolving systems, to the class of open harmonic quantum many-body systems. 
Since the {local Hilbert space dimension of bosonic degrees of freedom is not finite, the theorems of Secs.~\ref{extensions0} and \ref{extensions} do not apply.} However, locality bounds similar in mindset can still be derived. 
We again consider finite lattices $\Lambda$, equipped with a metric $d$, 
 {where now to each lattice site} $x\in \Lambda$ a Hilbert space ${\cal L}^2(\RR)$ is assigned, corresponding to a single harmonic mode with canonical coordinates $\{Q_x\}$ and $\{P_x\}$ 
 that satisfy the canonical commutation relations.
 The canonical coordinates in the lattice system can be  collected in the vector 
\begin{equation}
R=(Q_1,\dots, Q_{n}, 
Px_1,\dots, P_{n})
\end{equation}
with $n:= |\Lambda|$.
For clarity and ease of presentation, we formulate the locality bounds for commutators of canonical coordinates at different lattice sites, and for time-independent Liouvillians,
as analogs of the results of Section \ref{extensions0}. The results and techniques can however be extended to more general local observables, and also to time-dependent Liouvillians.

We consider the Gaussian setting in which the Lindblad operators {$L_\nu$} associated with each site $v\in \Lambda$ are linear in the canonical coordinates,
\begin{equation}\label{HLindblad}
	L_v = \sum_{j=1}^{2n} M_{v,j} R_j.
\end{equation}
We consider only a single Lindblad operator per site, although once again this assumption is not necessary and could be lifted.
The Hamiltonian is assumed to be a quadratic expression of the form
\begin{equation}\label{e:h_harmonic_gaussian}
H =\frac{1}{2}\sum_{x,y\in \Lambda} \left(Q_x A_{x,y} Q_y +
P_x B_{x,y} P_y
\right)
= \frac{1}{2}\sum_{x,y\in \Lambda} \left(R_x h_{x,y} R_y 
\right)
\end{equation}
with $h = A\oplus B$. In analogy with the discussion for spin systems in Section \ref{locality}, the Hamiltonian is said to be long-ranged {if the condition}
\begin{equation}\label{hpower}
|A_{x,y}|,|B_{x,y}|\leq \frac{c_0}{[1+d(x,y)]^\eta}
\end{equation}
is satisfied for positive constants $c_0,\eta$. As in Section ~\ref{extensions0}, we will require Assumption 1 from Section \ref{s:setting}, and are therefore restricted to $\eta>D$.
The Lindblad operators are required to obey the same spatial decay, 
\begin{equation}\label{mpower}
|M_{x,y}|,|M_{x,y+n}|\leq \frac{c_0}{[1+d(x,y)]^\eta}
\end{equation}
for all $x,y\in \Lambda$.
For this class of problems, the following Lieb-Robinson bound can be formulated:
\newline

\noindent\textbf{Theorem 4 (Harmonic Lieb-Robinson bound):} \textit{Given a finite lattice $\Lambda$, equipped with a metric $d$, and a
harmonic open quantum system described by a Liouvillian in the form of \eqref{HLindblad} and \eqref{e:h_harmonic_gaussian}, with the property that there exist positive constants $c_0$ and $\eta$ such that \eqref{hpower} and \eqref{mpower} are satisfied for all $x,y \in \Lambda$, and if Assumption 1 is satisfied for the same value of $\eta$, then for all $x,y\in\Lambda$, with $x\neq y$, and for all $0\leq s \leq t$, one has}
\begin{equation}
   \|[Q_x(s) ,Q_y]\|,\
		\|[P_x(s) ,P_y]\|,\,
	\|[P_x(s) ,Q_y]\|,\,
	\|[Q_x(s) ,P_y]\| \leq
	\frac{1}{2p_0}
    \frac{e^{2p_0 (c_0+ p_0 c_0^2)(t-s)}}{[1+d(x,y)]^\eta}.
\end{equation}

While Theorem 4 is stated for commutators of canonical coordinates and for time-independent Liouvillians, it should be clear from the proof below, in conjunction with the previously utilized proof techniques, how these assumptions can be lifted. Additionally, as in Theorem 2, the constraint of $\eta > D$ could be removed via a suitable rescaling of time facilitated by replacing Assumption 1 with Assumption 2. \newline

\proof
We start from the Heisenberg equations of motion \eqref{HE1} for the canonical coordinates,
\begin{equation} \label{e:Rk_Master}
\frac{d}{ds}R_k(s) = -\tilde{\mathcal{L}}^{\dagger}(s)R_k(s)
\end{equation}
for $k=1,\dots, 2n$, with $\tilde{\mathcal{L}}^{\dagger}$ acting as specified in Eq. \eqref{HE2}. By making use of the canonical commutation relations and the explicit form of $\tilde{\mathcal{L}}^{\dagger}$ specified through \eqref{HLindblad} and \eqref{e:h_harmonic_gaussian}, one finds
\begin{subequations}
\begin{align}
\frac{d}{ds}Q_x(s) 
&= - \sum_{y}B_{x,y} P_y (s)
- \frac{i}{2} \sum_{v\in \Lambda}   \sum_{j,l=1}^{2n}
	\left(\bar{M}_{v,j} M_{v,x} R_j(s) -
	\bar{M}_{v,x} M_{v,l} 
	R_l (s)
	\right),\\
\frac{d}{ds}P_x(s) 
&=  \sum_{y}A_{x,y} Q_y (s)
+ \frac{i}{2} \sum_{v\in \Lambda}   \sum_{j,l=1}^{2n}
	\left(\bar{M}_{v,j} M_{v,x} R_j(s) -
	\bar{M}_{v,x} M_{v,l} 
	R_l (s)
	\right).
\end{align}
\end{subequations}
Defining the matrices
\begin{subequations}
\begin{eqnarray}
	D_{x,y} &=& -\frac{i}{2} 
	\sum_{v=1}^n\bar{M}_{v,y} M_{v,x},\\
	E_{x,y} &=& 
	-\frac{i}{2} 
	\sum_{v=1}^n 
    \bar{M}_{v,y+n} M_{v,x},\\	
	F_{x,y} &=& \frac{i}{2} \sum_{v=1}^n 
	\bar{M}_{v,y} M_{v,x},\\
	G_{x,y} &=& \frac{i}{2} \sum_{v=1}^n 
	\bar{M}_{v,x} M_{v,y+n},
\end{eqnarray}
\end{subequations}
one can concisely write the equations of motion as
\begin{equation}\label{e:Rk_eom}
\frac{d}{ds}R_k(s) = \sum_{l=1}^{2n}S_{k,l}  R_l(s)
\end{equation}
with the kernel matrix
\begin{equation}
	S = \begin{bmatrix}
	0 & -B\\
	A & 0\\
	\end{bmatrix}
	+
	\begin{bmatrix}
	D &E\\
	-D & -E\\
	\end{bmatrix}
	+
	\begin{bmatrix}
	F &G\\
	-F & -G\\
	\end{bmatrix}
	=:\begin{bmatrix}
	S_1 & S_2\\
	S_3 & S_4
	\end{bmatrix}
	.\label{partition}
\end{equation}
The solutions to the equations of motion \eqref{e:Rk_eom} are given by the matrix exponential
\begin{equation}
	R_k(s) = \sum_{l = 1}^{2n}\big[e^{S(t-s)}\big]_{k,l} R_l.
\end{equation}
Note that the Hamiltonian part of this equation of motion is compatible with the findings of Ref.~\cite{HarmonicLR}. 
Following the strategy of Ref.\ \cite{HarmonicLR}, the Lieb-Robinson bounds of Theorem 4 then follow from an analysis of the matrix exponential. To this end {we upper bound the entries of the sub-matrices $S_1,\dots, S_4$ of $S$ defined in \eqref{partition}, for which from  the long-range conditions \eqref{hpower} and \eqref{mpower} it follows that}
\begin{equation}
	|(S_a)_{x,y}|\leq \frac{c_0}{[1+d(x,y)]^\eta}
	\sum_{v\in \Lambda\backslash\{x,y\}}
	\frac{c_0^2}{[1+d(x,v)]^\eta [1+d(y,v)]^\eta}
\end{equation}
for all $x,y\in \Lambda$ and $a=1,\dots, 4$. Making use of Assumption 1, we obtain the upper bound 
\begin{eqnarray}
	|(S_a)_{x,y}|&\leq &\frac{c_0+ p_0 c_0^2}{[1+d(x,y)]^\eta}.
\end{eqnarray}
In order to bound the matrix exponential, 
we need to bound the entries of powers of $S_a$. 
Invoking Assumption 1, we find
\begin{equation}
    ((S_a)^2 )_{x,y}  \leq p_0 (c_0+ p_0 c_0^2)^2
    \frac{1}{[1+d(x,y)]^\eta}.
\end{equation}
Iterating and generalizing this argument, one can show that
\begin{equation}
    {(S_{i_1}\cdots S_{i_k})_{x,y}}  \leq  p_0^{k-1} (c_0+ p_0 c_0^2)^k
    \frac{1}{[1+d(x,y)]^\eta}
\end{equation}
for any $i_1,\dots, i_k\in \{1,\dots, 4\}$. As a result, we find that for $x,y\in \Lambda$ (i.e. for the upper left  block of $S$)
\begin{equation}
    (S^k )_{x,y} \leq 2^{k-1} 
    p_0^{k-1} (c_0+ p_0 c_0^2)^k
    \frac{1}{[1+d(x,y)]^\eta},
\end{equation}
and the same inequality holds for the other three $n\times n$ blocks of $S^k$. 
This allows us to bound the entries of the matrix exponential, 
\begin{eqnarray}
    \big[e^{S(t-s)}\big]_{x,y}  &\leq & \left(\delta_{x,y}+
    \sum_{k=1}^\infty \frac{2^{k-1} }{k!}
    p_0^{k-1} (c_0+ p_0 c_0^2)^k (t-s)^k\right)
    \frac{1}{[1+d(x,y)]^\eta}\nonumber\\
    &= &\left(\delta_{x,y}+\frac{1}{2p_0}
    \sum_{k=1}^\infty \frac{2^{k} }{k!}
    p_0^{k} (c_0+ p_0 c_0^2)^k (t-s)^k\right)
    \frac{1}{[1+d(x,y)]^\eta}\nonumber\\
    &\leq &
    \left(\delta_{x,y}+\frac{1}{2p_0}
    e^{2p_0 (c_0+ p_0 c_0^2)(t-s)}\right)
    \frac{1}{[1+d(x,y)]^\eta}.
\end{eqnarray}
Theorem 4 then follows from application of the canonical commutation relations. \proofend\newline

\section{Summary}
In this work, we have presented a number of Lieb-Robinson bounds that
capture the locality of dynamics in long-ranged open quantum systems.
Such systems are currently a focus of interest, due to the fact that
several platforms for quantum simulators can be described well
by open systems of this type
\cite{FossFeig_etal13,Bohnet_etal16,Shankar_etal17,TrautmannHauke18,Malossi_etal14,Schempp_etal15,Zeiher_etal16,SchoenleberBentleyEisfeld18,Browaeys,Ott_etal13,GuerinAraujoKaiser16,Pucci_etal17}. 
It is the hope that this works stimulates
further research into the static and dynamical properties of such
quantum systems, e.g., by showing stability statements
\cite{Stability,Stability2} that follow 
from Lieb-Robinson bounds of the type we have presented here, or
to relate the findings established here
to experimental observations of open long-ranged
interacting systems out of equilibrium.

\begin{acknowledgments}
The authors gratefully acknowledge helpful discussions with David Storch and Michael Foss-Feig. R.~S.\ acknowledges the financial support of the Alexander von Humboldt foundation. J.~E.\ acknowledges financial support by the 
DFG (CRC 183 project B2, 
EI 519/7-1, FOR 2724).  This work has also received funding from the European Union's Horizon 2020	research and innovation programme 
under grant agreement No 817482 (PASQuanS).
M.~K.\ acknowledges financial support from the National Research Foundation of South Africa through the Competitive Programme for Rated Researchers.

\end{acknowledgments}	

\bibliography{main.bbl}

\end{document}